\documentclass[12pt]{article}
\usepackage{graphicx}

\newcommand\pubnumber{NuPhys2015-DiIura}
\newcommand\pubdate{}

\def\RomaTre{Dipartimento di Matematica e Fisica\\
Universit\`{a} di Roma Tre \textit{\&} INFN\\
Via della Vasca Navale 84, I-00146 Rome, ITALY}
\def\support{}

\def\Title#1{\begin{center} {\Large #1 } \end{center}}
\def\Author#1{\begin{center}{ \sc #1} \end{center}}
\def\Address#1{\begin{center}{ \it #1} \end{center}}

\newcommand\pubblock{\rightline{\begin{tabular}{l} \pubnumber\\
         \pubdate  \end{tabular}}}
\newenvironment{Abstract}{\begin{quotation}  }{\end{quotation}}
\newenvironment{Presented}{\begin{quotation} \begin{center} 
             PRESENTED AT\end{center}\bigskip 
      \begin{center}\begin{large}}{\end{large}\end{center} \end{quotation}}
\def\Acknowledgements{\bigskip  \bigskip \begin{center} \begin{large}
             \bf ACKNOWLEDGEMENTS \end{large}\end{center}}

\def\beq{\begin{equation}}
\def\eeq#1{\label{#1}\end{equation}}
\def\eeqn{\end{equation}}

\def\beqa{\begin{eqnarray}}
\def\eeqa#1{\label{#1}\end{eqnarray}}
\def\eeqan{\end{eqnarray}}

\let\bar=\overbar

\def\Dslash{\not{\hbox{\kern-4pt $D$}}}
\def\dslash{\not{\hbox{\kern-2pt $\del$}}}

\def\msb{{\bar{\ssstyle M \kern -1pt S}}}

\usepackage{bbm}

\newcommand{\Gr}{\mathcal{G}}
\newcommand{\Id}{\mathbbm{1}}
\newcommand{\mbf}{\mathbf}
\newcommand{\UPMNS}{U_{\mathrm{PMNS}}}
\newcommand{\chisquare}{\chi^2}
\newcommand{\chimin}{\chi^2_{\mathrm{min}}}
\newcommand{\thbf}{\theta_{\mathrm{bf}}}

\begin{document}
\begin{titlepage}
\pubblock

\vfill
\Title{Lepton mixing and neutrino masses from $A_5$ and $CP$}
\vfill
\Author{ Andrea Di Iura\support}
\Address{\RomaTre}
\vfill
\begin{Abstract}
Some properties of lepton mixing and neutrino masses can be computed under the assumption of $A_5$ and $CP$ as a symmetry in the leptonic sector. The results show that four mixing patterns accommodate well the oscillation data, i.e. all the mixing angles are in the $3\sigma$ confidence region. We also introduce an explicit realization of this framework in the case of the Weinberg operator where the neutrino mass spectrum can be computed. 
\end{Abstract}
\vfill
\begin{Presented}

NuPhys2015, Prospects in Neutrino Physics

Barbican Centre, London, UK,  December 16--18, 2015

\end{Presented}
\vfill
\end{titlepage}
\def\thefootnote{\fnsymbol{footnote}}
\setcounter{footnote}{0}

\section{Mixing patterns}

\begin{figure}[h!]
\begin{center}
\includegraphics[height=1.9in]{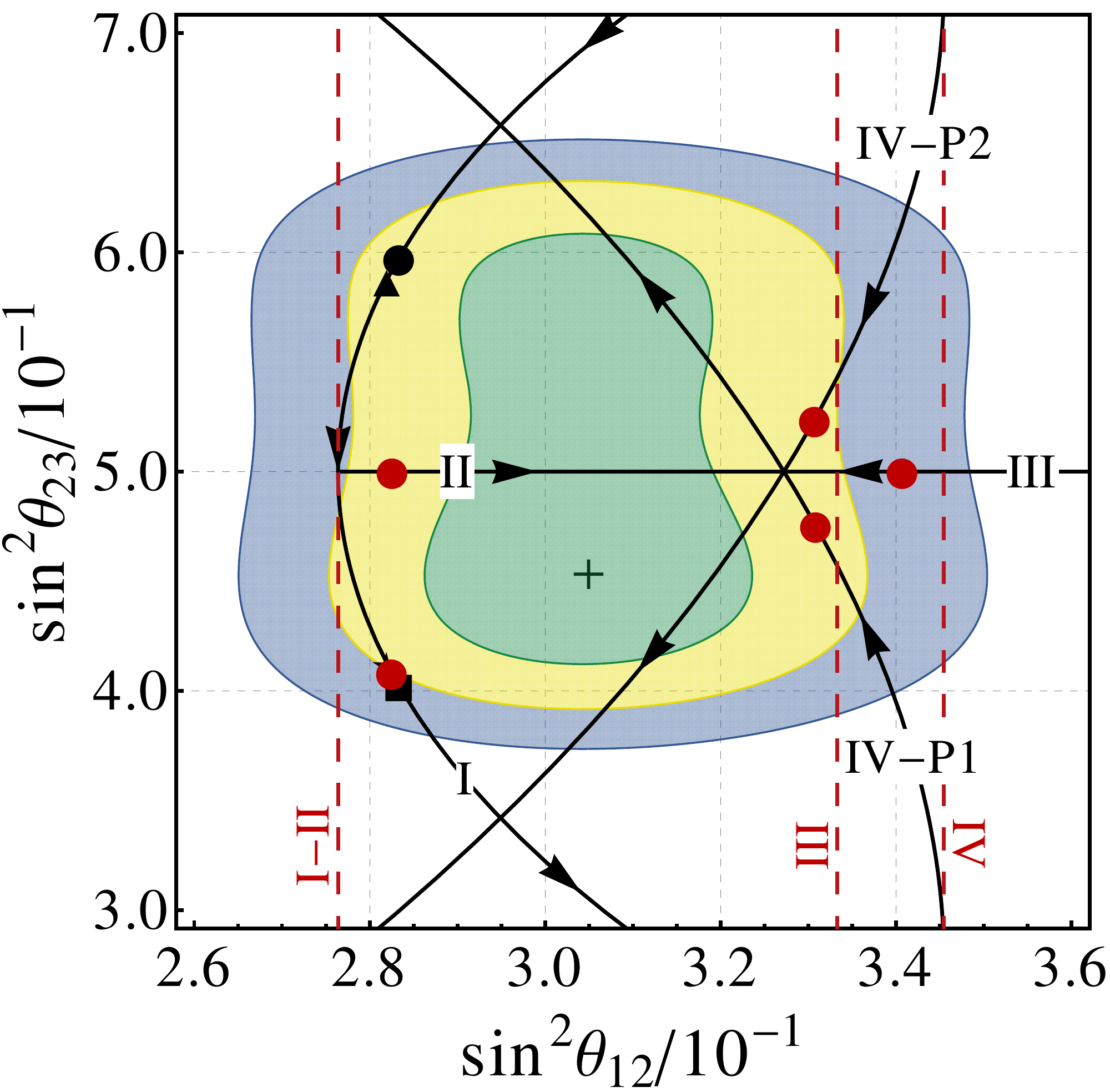}
\includegraphics[height=1.9in]{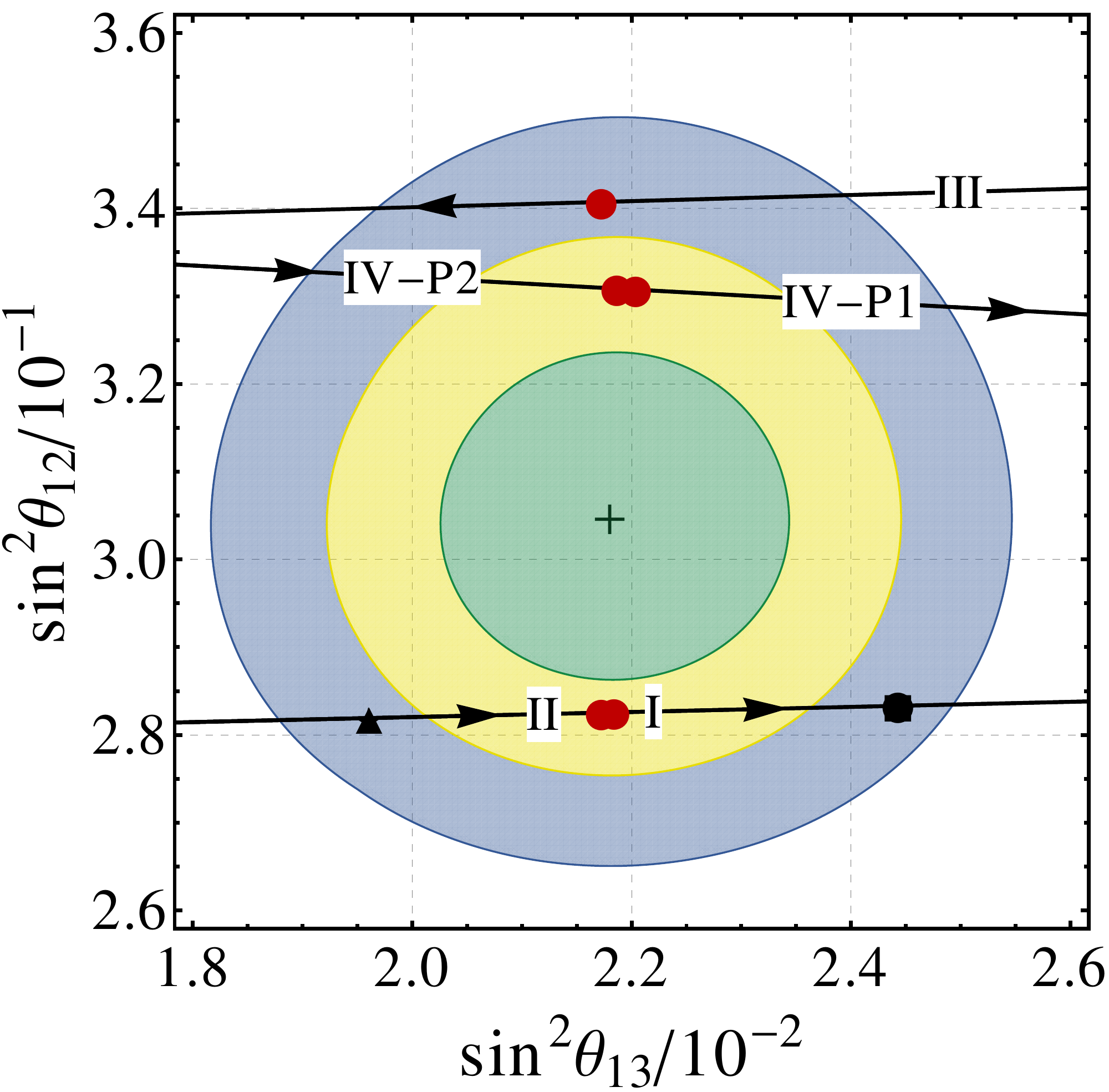}
\includegraphics[height=1.9in]{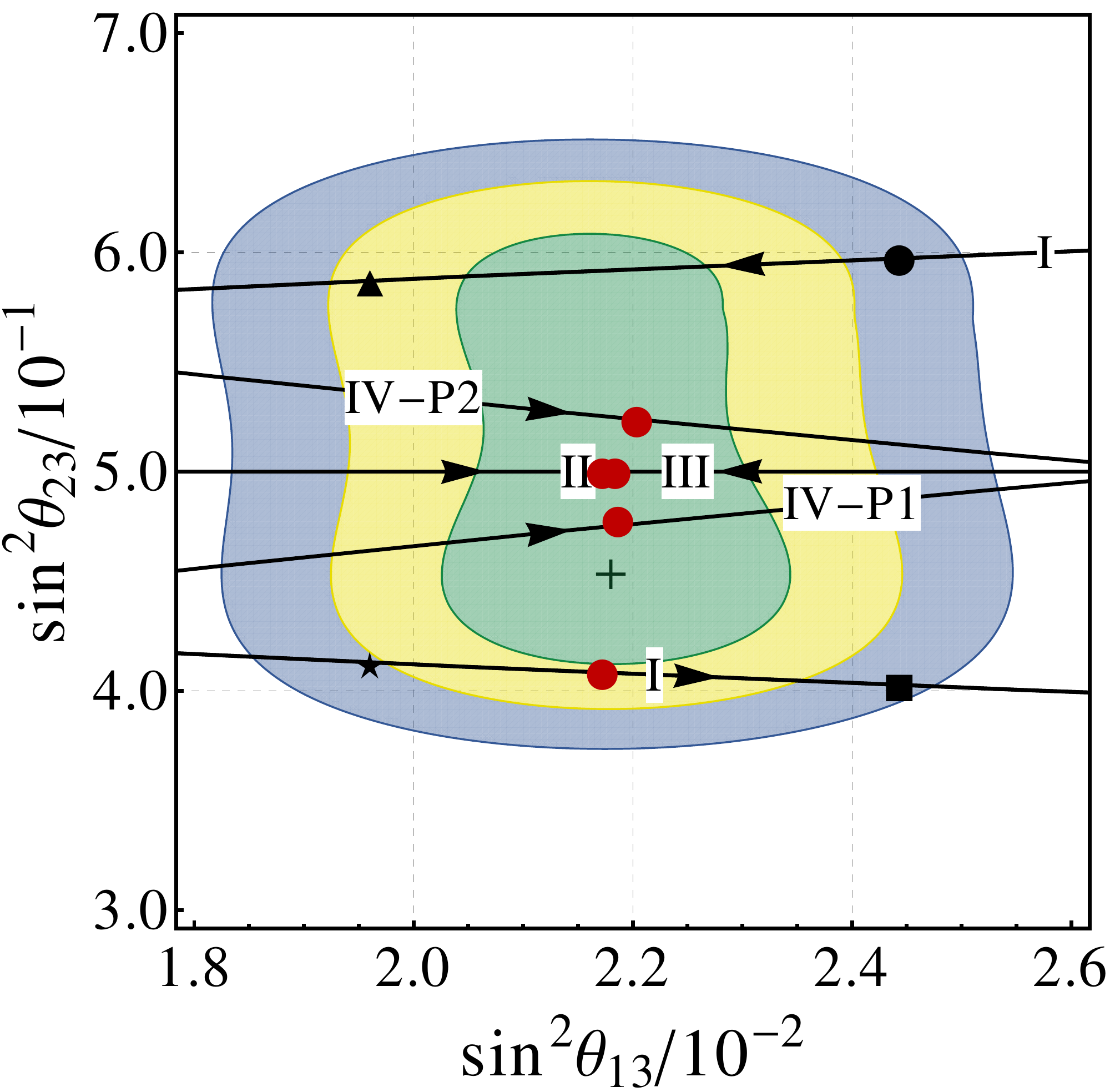}
\caption{Parametric plots for Case I trhought Case IV as a function of the internal angle $\theta$ in the planes $(\sin^2\theta_{12}, \sin^2\theta_{23})$ (left), $(\sin^2\theta_{13}, \sin^2\theta_{12})$ (middle) and $(\sin^2\theta_{13}, \sin^2\theta_{23})$ (right) assuming NO. The red dots are the model best fit values while the black crosses are the experimental best fit points. The shadow areas are the allowed regions at $1\sigma$, $2\sigma$ and $3\sigma$ CL (2dof).}
\label{fig:patterns_A5CP}
\end{center}
\end{figure}

A possible tool to understand the mixing pattern in the leptonic sector that we observe in Nature is the one based on Flavour Symmetry. In particular in our approach we use non abelian discrete symmetry combined with $CP$ as discussed in \cite{FERUGLIO_HAGEDORN_ZIEGLER, HOLTHAUSEN_LINDNER_SCHMIDT}. In our study, fully presented in \cite{DIIURA_HAGEDORN_MELONI}, we assume that the group of full symmetry in the leptonic sector is $\Gr_f = A_5 \otimes CP$, where $A_5$ is the group of even permutation of five elements. Neutrinos are Majorana particles in representation $\mbf{3} \in A_5$. The generators' algebra is
\begin{equation}
S^2 = T^5 = (ST)^3 = \Id.
\end{equation}
The PMNS matrix is given by the misalignment between the residual symmetries in charged and neutrino sector, respectively $\Gr_{\ell}$ and $\Gr_{\nu}$. For the charged leptons we consider $\Gr_{\ell} = \{Z_3, Z_5, Z_2 \otimes Z_2\}$ abelian subgroups of $A_5$, and $Q$ is a representation of the subgroup in the field space, $Q \in Z_m \Longleftrightarrow Q^{m} = \Id$. As a residual symmetry in the neutrino sector we assume $\Gr_{\nu}= Z_2 \otimes CP$ and we define $Z \in Z_2$ and $X$ is a representation of $CP$ in the field space such that $X X^T = X X^\star = \Id$, $X Z^\star -ZX = 0$ and it is an involutive automorphism of the flavour group $\Gr_f$, as discussed in details in \cite{CHEN_ET_AL_CP}.\\
In the following $X_0$ is the simplest representation of $CP$ symmetry. It is the permutation matrix in the $2-3$ plane. The other representations of $CP$ can be obtained as $X = ZX_0$. The action of the symmetry on charged lepton mass matrix $M_{\ell}$ and neutrino mass matrix $M_{\nu}$ is the following
\begin{equation}
 Q^\dagger M_\ell^\dagger M_\ell Q=M_\ell^\dagger M_\ell \quad Z^T M_\nu Z = M_\nu \quad X M_\nu X = M_\nu^\star.
\end{equation}
For each possible touple $(Q, Z, X)$ the PMNS matrix can be constructed as
\begin{equation}
\left\{\begin{array}{l}
 U_\ell^\dagger Q U_\ell = Q_{diag}\\
 \Omega^\dagger Z \Omega = Z_{diag}\\ X = \Omega \Omega^T
\end{array} \right)\Longrightarrow \UPMNS = U_\ell^\dagger U_\nu = U_\ell^\dagger \Omega R_{ij}(\theta) K_\nu \quad \theta \in [0, \pi)
\end{equation}
where the rotation matrix $R_{ij}(\theta)$ is necessary to correctly diagonalize the neutrino mass matrix $M_\nu$, and $K_\nu$ is a diagonal matrix needed to have all neutrino masses positive. Therefore all the oscillation parameters are function of the internal angle $\theta$. The power of this approach is that all the $CP$ phases can be predicted.\\
For each choice of the tuple $(Q, Z, X)$ we have a PMNS matrix up to permutations of rows and columns since the masses are not fixed in this approach. The independent tuples are 18. We use a $\chisquare$ function, based on the data reported in \cite{GONZALEZGARCIA_ET_AL}, in order to reduce the number of independent tuple to four, which are relevant for the phenomenology, see Table~\ref{tab:pattern} and Figure~\ref{fig:patterns_A5CP}. Similar results were obtained in \cite{DING_A5, PASCOLI_A5}.

\begin{table}[h!]
\begin{center}
\begin{tabular}{c  c c   c c c  c }
\bf Case & $\chimin$ & $\thbf$ & $\sin^2\theta_{12}$ & $\sin^2\theta_{13}$ & $\sin^2\theta_{23}$ &  $\sin\delta$\\ 
\hline
\bf$\Gr_\ell = Z_5$ - Case I &  			5.64 	& 0.174  & 0.283 & 0.0217 & 0.408  & 0\\
$(T^2, T^2ST^3ST^2, SX_0)$& 	3.46 	& 2.967  & 0.283 & 0.0219 & 0.592  & 0\\
\hline
\bf $\Gr_\ell = Z_5$ - Case II &  			4.04 	& 0.175-2.967  & 0.283 & 0.0218 & 0.5 & $\mp$1\\
$(T^2, ST^2ST, X_0)$& 		7.74  	& 0.175-2.967  & 0.283 & 0.0220 & 0.5 & $\mp$1\\
\hline
\bf $\Gr_\ell = Z_3$ - Case III &  		8.84 	& 0.604-0.967  & 0.341 & 0.0217 & 0.5 & $\pm$1\\
$(T^2ST^2, ST^2ST^3S, X_0)$& 	12.56  	& 0.603-0.967  & 0.341 & 0.0218 & 0.5 & $\pm$1\\
\hline
\bf $\Gr_\ell = Z_2\otimes Z_2$ - Case IV-P1 &  		4.48  	& 0.254  & 0.331 & 0.0219 & 0.475 & 0\\
$(\{ S, T^2 S T^3 S T^2 \} , S T^2 S T , X_0)$& 	11.80  	& 0.258  & 0.330 & 0.0225 & 0.478 & 0\\
\hline
\bf $\Gr_\ell = Z_2\otimes Z_2$ - Case IV-P2 &  		6.19   	& 0.255  & 0.331 & 0.0220 & 0.524 & 0\\
$(\{ S, T^2 S T^3 S T^2 \} , S T^2 S T , X_0)$& 	6.43  	& 0.254  & 0.331 & 0.0218 & 0.525 & 0\\
\hline
\end{tabular}
\caption{Values of $\chimin$, best fit for $\theta$ and PMNS parameters for patterns that have $\chimin \leq 27$. Upper rows are for Normal Ordering (NO) while lower ones are for Inverted Ordering (IO). Notice that the Dirac phases are maximal when also the atmospheric angles are maximal, otherwise $\delta$ is trivial. The Majorana phases are always trivial if we want to accommodate well the mixing angles, but patterns with non trivial values can appear.}
\label{tab:pattern}
\end{center}
\end{table}

\section{Explicit Model: Case II}

\begin{figure}[htb]
\begin{center}
\includegraphics[height=2in]{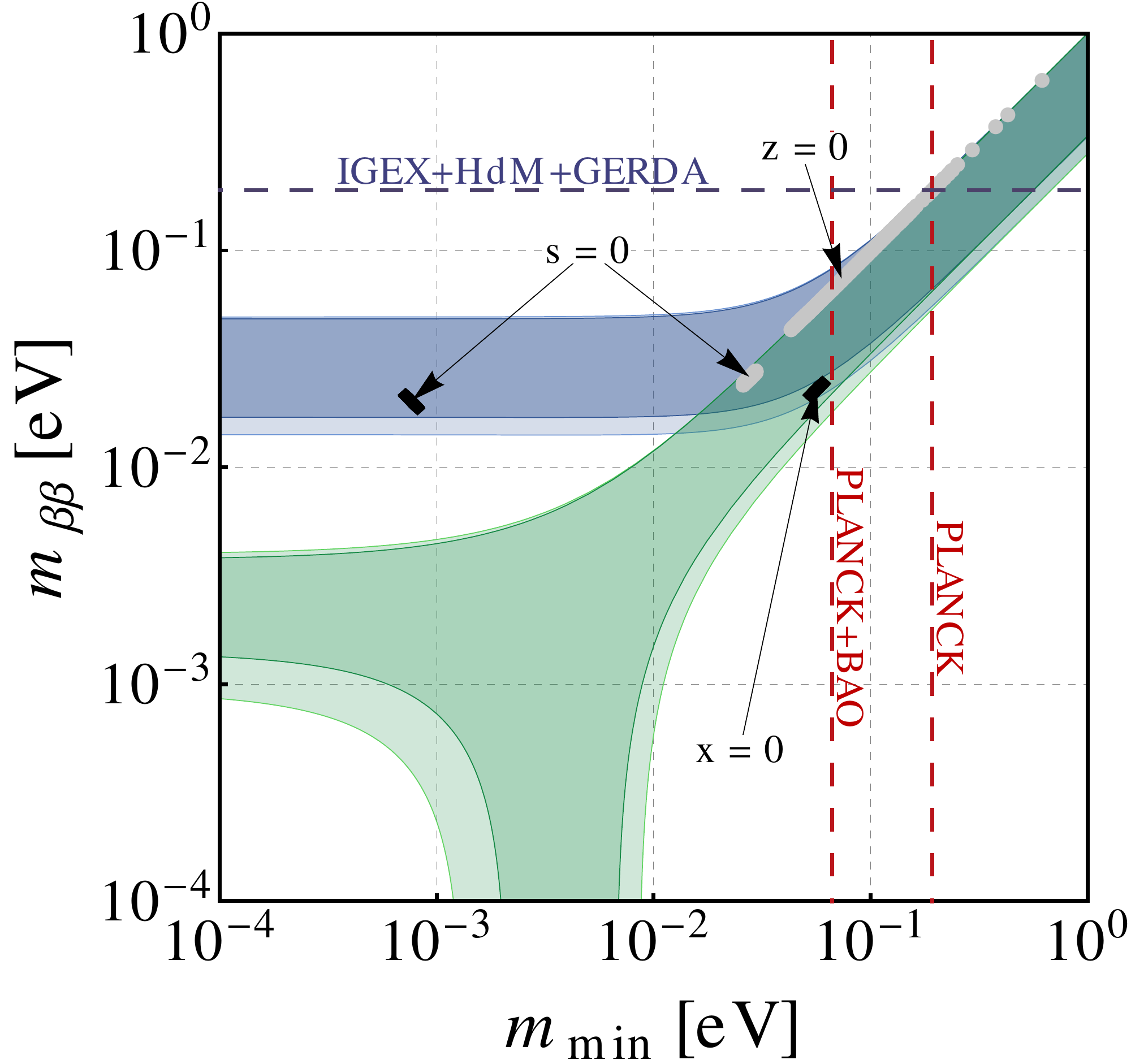}
\includegraphics[height=2in]{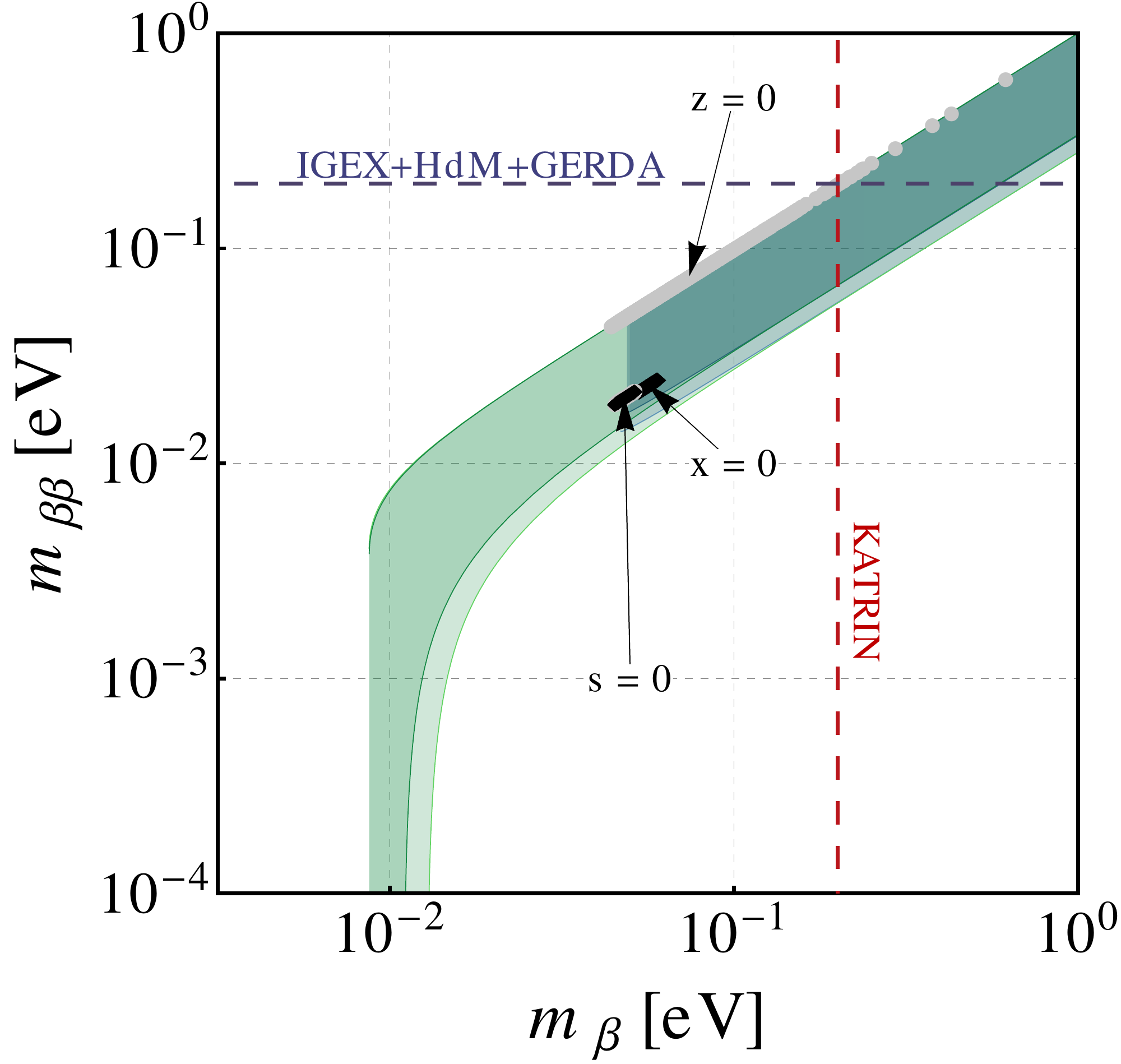}
\caption{Predictions for the $0\nu\beta\beta$-decay effective mass $m_{\beta \beta}$ as function of the lightest neutrino mass $m_{\mathrm{min}}$ (left) or $m_{\beta}$ (right) in the case of the Weinberg operator assuming one vanishing parameter in the neutrino mass matrix $M_\nu$. The Majorana phases are fixed in explicit models. Gray circles are for NO while black diamonds for IO.}
\label{fig:caseII_A5CP}
\end{center}
\end{figure}

The mixing patterns are independent on the underlying theory, they are just a consequence of group theory. However if we want to obtain other informations about neutrino physics we need to construct an explicit model. We focus on Case II of our classification, which has $Z_5$ as a residual symmetry for the charged lepton sector. The oscillation parameters are
\begin{equation}
 \sin^2\theta_{13} = \frac{\varphi + 2}{5}\sin^2\theta \quad \sin^2\theta_{12} = \frac{3-\varphi}{5\cos^2\theta_{13}} \quad \sin^2\theta_{23} = \frac{1}{2}\quad |\sin\delta| = \pm 1
\end{equation}
where $\varphi \equiv (1+\sqrt{5})/2$ is the Golden Ratio. The neutrino mass matrix $M_\nu$ is fixed by symmetry. $M_\nu$ has four real dimensionless parameters $s, x, y, z$ and a scale factor $m_0$. The internal angle $\theta$ is related to the $M_\nu$ parameters
\begin{equation}
  \frac{M_\nu}{m_0} = \left(\begin{array}{c c c}
          s + x + z & \frac{3}{2\sqrt{2}}(z + i \varphi y) & \frac{3}{2\sqrt{2}}(z - i \varphi y)\\
          \frac{3}{2\sqrt{2}}(z + i \varphi y) & \frac{3}{2} (x + i y) & s - \frac{x + z}{2} \\
          \frac{3}{2\sqrt{2}}(z - i \varphi y)  & s - \frac{x + z}{2} & \frac{3}{2} (x - i y) 
         \end{array}\right)  \ \tan 2 \theta = \frac{2\sqrt{7 + 11 \varphi}y}{2 x(\varphi +1) + z(2\varphi + 1)}.
\end{equation}
 A small value of $|y|$ is needed to obtain $\theta_{13} \sim 9^\circ$. Assuming a Weinberg operator with two flavon fields $\phi_{\nu, \mbf{1}} \sim \mbf{1} \in A_5$ and $\phi_{\nu, \mbf{5}} \sim \mbf{5} \in A_5$ is possible to achieve a small reactor angle in a two step symmetry breaking $A_5 \otimes CP \to Z_2 \otimes Z_2 \otimes CP \to Z_2 \otimes CP$ because under the Klein group and $CP$ the parameter $y$ is vanishing and only under $Z_2 \otimes CP$ it appears. Therefore $y$ is naturally the smallest parameter.\\
The parameter space with four free dimensionless parameters is sufficiently large to describe all the oscillation data, thus we start a classification reducing the number of independent VEVs in the flavon potential to obtain predictive scenarios \cite{DIIURA_HAGEDORN_MELONI_2}. In these limits we can predict the mass spectrum and obtain information about the Majorana phases.  For instance, in Figure~\ref{fig:caseII_A5CP}, the $0\nu\beta\beta$-decay effective mass is shown as function of the lightest neutrino mass or the effective $\beta$-decay mass. 

\section{Conclusion}

The use of non abelian discrete symmetry and $CP$ is a useful approach to understand the leptonic mixing pattern, in particular the smallness of $\theta_{13}$. Assuming $A_5 \otimes CP $ as a global symmetry in the full leptonic sector and $Z_2 \otimes CP$ as a residual symmetry in the neutrino sector we obtain that four mixing patterns are relevant for the phenomenology.\\
We also study an explicit realization of this framework based on Case II which has $Z_5$ as a residual symmetry in the charged sector. In our classification we reduce the number of independent VEVs obtaining predictive scenarios with only three free parameters because the Majorana phases are fixed. We observed that in these limits interesting feature appears, for instance in the case of Weinberg operator with $z = 0$ where the ratio between the solar and atmospheric mass splittings is proportional to $\sin^2\theta_{13}$.

\Acknowledgements
I gratefully acknowledge C. Hagedorn and D. Meloni for the fundamental help in the development of our work.

\end{document}